\documentclass[12pt,a4paper,dvips]{article}
\usepackage{epsfig}
\usepackage{a4p}
\usepackage{cite,mcite}
\usepackage{graphicx}
\usepackage{latexsym}
\usepackage{amssymb}
\usepackage{physics}
\usepackage{l3_title,ifthen,Lep}
\preprint{2000-057}
\date{April 28, 2000}
\journalname{Phys. Lett. B}
\def\ZG{\ensuremath{\Zo\gamma}}

\def\EEZG{\ensuremath{\ee\ra\ZG}}

\def\QQG{\ensuremath{\qqbar\gamma}}
\def\NNG{\ensuremath{\nnbar\gamma}}
\def\EEQQG{\ensuremath{\ee\ra\qqbar\gamma}}
\def\EENNG{\ensuremath{\ee\ra\nnbar\gamma}}
\def\ZZG{\ensuremath{\Zo\Zo\gamma}}
\def\ZGG{\ensuremath{\Zo\gamma\gamma}}
\def\ZVG{\ensuremath{\Zo V\gamma}}
\def\THFCM{\ensuremath{\theta_f^\Zo}}
\def\PHFCM{\ensuremath{\phi_f^\Zo}}

\newlength\leftl
\newlength\downl
\newlength\rightl
\newlength\templ
\def\spb#1#2#3#4#5{%
  \setbox0\hbox{#1}\setbox1\hbox{\char'023}%
  \rightl=#2\wd0 \advance\rightl by-#3\wd1
  \downl=#5\ht1 \advance\downl by-#4\ht0
  \leftl=\rightl \advance\leftl by\wd1
  \ht1=\downl \dp1=-\downl
  \leavevmode
  \kern\rightl\lower\downl\box1\kern-\leftl #1}
\def\lpb{\spb l{.35}{.5}{.4}{.8}}%
\newlength{\capindent}
\newlength{\capwidth}
\newlength{\figwidth}
\newcommand{\icaption}[2][!*!,!]{\hspace*{\capindent}%
  \begin{minipage}{\capwidth}
    \ifthenelse{\equal{#1}{!*!,!}}%
      {\caption{#2}}%
      {\caption[#1]{#2}}
  \end{minipage}}
\begin{document}
\begin{titlepage}
\title{Search for Anomalous \mbox{\boldmath $\ZZG$} and \mbox{\boldmath $\ZGG$} 
couplings in the process \mbox{\boldmath $\epem \ra \ZG$} at LEP} 
\author{The L3 Collaboration}
%
%
\begin{abstract}
\indent
We search for anomalous trilinear gauge couplings in the $\Zo\Zo\gamma$ 
and $\Zo\gamma\gamma$ vertices using data collected with the L3 detector
at LEP at a centre--of--mass energy $\sqrt{s}=189\GeV$. No evidence is 
found and limits on these couplings and on new 
physics scales are derived from the analysis of the process 
$\epem \ra \Zo \gamma$.
\end{abstract}
\submitted
\end{titlepage}

%
%
\section*{Introduction}

The process $\epem \ra \Zo \gamma$ is interesting to test the existence
of new physics\cite{renard}. In particular, the existence of anomalous couplings
between neutral gauge bosons can be probed by means of this reaction. Effects 
coming from $\Zo\Zo\gamma$ and $\Zo\gamma\gamma$ couplings are expected to 
be very small in the Standard Model\cite{renard,barroso}, but can be enhanced 
in compositeness models\cite{renard2,SCSM} or if new particles enter in higher 
order corrections. 

Assuming only Lorentz and U(1)$_{em}$ gauge invariance, the most general 
form of the $\ZZG$ and $\ZGG$ vertices can be parametrized by means of
eight anomalous couplings. The couplings are: 
$h_i^V~(i=1~{\rm to}~4; V=\gamma,\Zo)$, where
a $V$ superscript corresponds to a $\ZVG$ vertex. The couplings 
$h_1^V$ and $h_2^V$ are CP violating whereas $h_3^V$ and $h_4^V$ are CP 
conserving. The anomalous couplings contribution to the $\ZG$ cross section
increases with the centre--of--mass energy, $\sqrt{s}$, while the Standard Model 
contribution decreases. All these couplings are zero at tree level in the
Standard Model, and only the CP conserving ones receive a small contribution
($\approx 10^{-4}$) at the one loop level{\protect \cite{renard,barroso}. An
alternative parametrization, which introduces the energy scales of
new physics $\Lambda_{iV}$ is \cite{mery}:

\begin{eqnarray}
   \frac{\sqrt{\alpha}~h_i^V}{\MZ^2} \equiv \frac{1}{\Lambda_{iV}^2} & , i=1,3
                                                            \label{for:lambda13} \\
   \frac{\sqrt{\alpha}~h_i^V}{\MZ^4} \equiv \frac{1}{\Lambda_{iV}^4} & , i=2,4.
                                                            \label{for:lambda24}
\end{eqnarray}

The anomalous couplings $h_i^V$ cannot get arbitrarily large values due to
unitarity constraints, and should vanish when $s \ra \infty$ 
\cite{polish,baur}. For hadron colliders, this means that a form factor
dependence should be used \cite{baur,cdf,d0,d0_new}, because the effective
centre--of--mass energy of the collision is variable. For $\epem$ colliders, the
centre--of--mass energy is fixed and no form factors are 
needed \cite{gounaris}. Therefore, the results are model independent and 
no hypothesis about the behaviour close to the new physics scale is
made \cite{wudka,gounaris}. The main effects of anomalous 
$\Zo\Zo\gamma$ and $\Zo\gamma\gamma$ couplings
are an increase of the $\Zo\gamma$ total cross section and a modification
of the differential spectrum of the photon, mainly at large polar 
angles. Limits on $\Zo\Zo\gamma$ and $\Zo\gamma\gamma$ couplings have
been published at the Tevatron\cite{cdf,d0,d0_new} and LEP \cite{us,delphi}.

Data collected by the L3 detector \cite{L3det} in 1998 at 
$\sqrt{s}=189\GeV$, amounting to 
176 pb$^{-1}$ of integrated luminosity, are used to study the anomalous 
couplings $\ZZG$ and $\ZGG$ in the channels $\EEQQG$ and \EENNG. In this 
article, a new calculation \cite{gounaris} is 
used.\footnote{Previous publications use reference \cite{hagiwara} as 
the theoretical input, however recently\cite{gounaris} an error in the 
formulation was found. Thus previous limits cannot be directly compared 
with those obtained in this paper.} 
The results presented in this paper 
supersede our previous results \cite{us} due to the higher $\sqrt{s}$ and 
integrated luminosity, which improve the sensitivity, and to the new 
calculation used.  

%
%
\section*{Event Selection}

The main signature of the process $\EEZG$ is the production of a photon of 
high energy, $E_{\gamma}$. A photon candidate is 
identified as a shower in the BGO electromagnetic barrel or end-cap calorimeters
with more than 90\% of its energy deposited in a $3 \times 3$ crystal matrix. The
mass of the system recoiling against the 
photon, $m_{rec}=(s-2E_{\gamma}\sqrt{s})^{1/2}$, is required to satisfy
$80\GeV < m_{rec} < 110\GeV$, and thus is consistent with that of a $\Zo$. For 
the $\sqrt{s}$ value considered, the cuts on the recoiling mass correspond to 
an energy of the photon between $62\GeV$ and $78\GeV$. 

In the estimation of signal and background processes the following Monte Carlo 
generators have been used: KK2F \cite{kk2f} for $\EEQQG$, $\epem \ra \qqbar$, KORALZ 
\cite{koralz} for $\EENNG$, EXCALIBUR \cite{EXCALIBUR} for four fermion final 
states, DIAG36 \cite{diag36} for $\ee \ra \ee \ee$, BHWIDE \cite{BHWIDE} and 
TEEGG \cite{TEEGG} for $\ee \ra \ee\gamma(\gamma)$. All generated events are passed 
through a simulation of the L3 detector response \cite{geant} and through the same 
analysis procedure used for data. Time dependent inefficiencies are taken into 
account in the simulation procedure.

Cross sections and efficiencies are quoted within the following phase 
space cuts: at least one photon of energy greater than $20 \GeV$ with polar 
angle in the range $5^\circ < \theta_\gamma < 175^\circ$.

%
%
\subsection*{Selection of  \mbox{\boldmath{$\EEQQG$}} events}

In addition to a photon candidate recoiling against a
$\Zo$, high multiplicity and energy-momentum balance are required 
to select $\EEQQG$ events:

\begin{itemize}
\item The event must have more than 6 tracks and more than 11 calorimetric
      clusters.

\item The transverse energy imbalance must be less than 15\% of the 
      visible energy, and the longitudinal energy imbalance less 
      than 20\% of the visible energy. 

\item Events are rejected if the photon candidate is isolated from the
      two jets coming from the $\Zo$ and if it is associated with a
      track. This requirement eliminates a substantial contamination of
      $\ee \ra {\rm q \bar{q}' e}\nu$ and $\ee \ra \qqbar \ee$ events.

\item The polar angle of the photon, $\theta_\gamma$, must be
      in the range $|\cos\theta_{\gamma}|< 0.74$ or $0.82<|\cos\theta_{\gamma}|<0.97$.
\end{itemize}

Using these criteria, 899 events are selected. The trigger inefficiency is
estimated to be negligible due to the redundancy of subtriggers involved in
triggering this final state.

Three backgrounds are found to contribute: $\ee \ra {\rm q \bar{q}' e}\nu$ and 
$\ee \ra \qqbar \ee$, where one of the electrons fakes a photon, giving 
$0.5 \%$ and $0.4 \%$ contamination respectively, and 
$\ee \ra \qqbar$ events, contributing to a $1.2 \%$ contamination, mainly 
due to photons from $\pio$. 

The measured cross section for $\EEQQG$ is shown in Table \ref{tab:eff_qqg} and is in
agreement with the Standard Model expectation. Figures \ref{fig:recomass_qq} and 
\ref{fig:thetafo_qq} show the recoiling mass distribution and the polar angle 
of the photon. 

\begin{table}[t]
\centering
\begin{tabular}{|c|c|c|c|c|c|}
\hline
  Process       & ${\cal L}$ ($\pb$) & $\epsilon$ (\%) & Events &
                                       $\sigma$ (pb)   & $\sigma_{\rm SM}$ (pb)\\
\hline
 $\EEQQG$ & 172.1 & 28.9 $\pm$ 0.1 & 899 & $           18.1 \pm 0.6$ &            $17.9$ \\ 
 $\EENNG$ & 175.6 & 33.6 $\pm$ 0.2 & 288 & $\phantom{0} 4.8 \pm 0.3$ & $\phantom{0} 5.0$ \\
\hline
\end{tabular}
\icaption{Integrated luminosities, ${\cal L}$, efficiencies, $\epsilon$, of the 
          selection, number of selected events and measured cross sections. The 
          error on the efficiencies accounts for Monte Carlo statistics. The 
          corresponding Standard Model cross sections $\sigma_{\rm SM}$ 
          {\protect \cite{kk2f,koralz}} are also listed.
\label{tab:eff_qqg}}
\end{table}

%
%
\subsection*{Selection of \mbox{\boldmath{$\EENNG$}} events}

In addition to the presence of a photon the selection criteria for the 
$\EENNG$ channel require low activity in the detector and large 
energy-momentum imbalance.

\begin{itemize}
\item The event cannot have more than 10 calorimetric clusters. The 
      number of hits in the tracking chamber associated to a
      calorimetric cluster must not exceed 40\% of the expected number of hits
      for a charged track.

\item The polar angle of the photon candidate must satisfy
      $|\cos\theta_{\gamma}|< 0.74$ or $0.82<|\cos\theta_{\gamma}|<0.96$.

\item The transverse and total energy imbalances in the event must be greater 
      than 20\% and 95\% of the visible energy, respectively.

\item To suppress cosmic ray background, there must be at least one scintillator 
      time measurement within $\pm 5$ ns of the beam crossing time. The scintillator 
      signals must be associated to calorimetric clusters.

\end{itemize}

A total of 288 events are selected. The trigger efficiency is estimated to 
be $93.5 \%$, using an independent sample of data events.

The background in the selected sample is found to be negligible. The measured 
cross section for $\EENNG$ is shown in Table \ref{tab:eff_qqg}. It is in agreement with 
the Standard Model expectation. Figure \ref{fig:recoilmass_nunug} shows 
the recoiling mass distribution and Figure \ref{fig:costheta_nunug} shows 
the polar angle of the photon.

%
%
\section*{Results}

The analysis is performed by means of the ``Optimal Variables'' \cite{oo} 
approach. The differential cross section of the process $\EEZG$ is written 
as a function of the anomalous couplings:

\begin{eqnarray}
\frac{d\sigma}{d\vec{\Omega}}=c_{\scriptscriptstyle 0}(\vec{\Omega})+
\sum^{4}_{i=1}\sum_{V=\gamma,\Zo} c_{\scriptscriptstyle 1,i}(\vec{\Omega}) h_{i}^{V}+
\sum^{4}_{i=1}\sum_{V=\gamma,\Zo}\sum^{4}_{j=1}\sum_{V'=\gamma,\Zo}c_{\scriptscriptstyle 2,ij}(\vec{\Omega})
h_{i}^{V}h_{j}^{V'},
\end{eqnarray}

\noindent
where $\vec{\Omega}=(E_\gamma, \theta_\gamma, \phi_\gamma, \THFCM, \PHFCM)$. $E_\gamma, \theta_\gamma$ 
and $\phi_\gamma$ are the energy and angular variables of the
photon, and $\THFCM, \PHFCM$ the angles of the fermion $f$ in the $\Zo$ rest 
frame. For the $\EENNG$ case, the neutrino angular 
variables are integrated out, leaving three variables in the calculation. As the
couplings are small, the quadratic term is neglected, and the 
optimal variable for each coupling is defined as: 

\begin{eqnarray}\phantom{0}
{\cal O}_{1,i} & \equiv & \frac{c_{\scriptscriptstyle 1,i}(\vec{\Omega})}
{c_{\scriptscriptstyle 0}(\vec{\Omega})}.
\end{eqnarray}

\noindent
This method profits from having a one-dimensional 
parametrization which uses all the available kinematic information.

A binned maximum likelihood fit of the expected optimal variable distribution
to the data is performed. A reweighting technique is used to compute the expected number
of events in the presence of anomalous couplings. In these fits, only one
coupling is fitted each time, keeping all the other couplings at 
zero. The distributions of the optimal variables for the couplings
$h_{1}^{\Zo}$ and $h_{4}^{\gamma}$ are shown as an example in Figures \ref{fig:h1z_oo} 
and \ref{fig:h4g_oo}, respectively. 

No deviations from the Standard Model expectations are found. Both $\QQG$ and 
$\NNG$ samples lead to the same results. The limits on the anomalous couplings
at 95\% confidence level (CL) coming from both samples are: 

\begin{center}
\begin{tabular}{lcl}
   $-0.26 < h_{1}^{\Zo}  < 0.09$ & ~~ & $-0.20 < h_{1}^\gamma < 0.08$   \\ 
   $-0.10 < h_{2}^{\Zo}  < 0.16$ & ~~ & $-0.11 < h_{2}^\gamma < 0.11$   \\ 
   $-0.26 < h_{3}^{\Zo}  < 0.21$ & ~~ & $-0.11 < h_{3}^\gamma < 0.03$   \\ 
   $-0.11 < h_{4}^{\Zo}  < 0.19$ & ~~ & $-0.02 < h_{4}^\gamma < 0.10$.  \\
\end{tabular}
\end{center}

An independent analysis using a maximum likelihood fit in the five-dimensional
phase space, described in our previous publication \cite{us} is also performed
as a cross check. The results obtained using this method are compatible with
those obtained with the optimal variables.

Fits to the two-dimensional distributions of the optimal observables are performed
to determine the pairs of the CP-violating and CP-conserving couplings, keeping 
in each case the other couplings fixed at zero. Results 
at $95\%$ CL are shown in Table \ref{tab:h_corr}. There exists a strong 
correlation between the fitted values of the CP-violating couplings ($h_1^V,h_2^V$) or 
CP-conserving couplings ($h_3^V,h_4^V$). Contours for $95\%$ CL 
two-dimensional limits on each pair of couplings, for $\ZZG$ and $\ZGG$, are shown in 
Figures \ref{fig:elipse2} and \ref{fig:elipse1}.

\begin{table}[htbp]
\centering
\begin{tabular}{|c|r|c|c|r|}
\hline
                &  Fitted   & Negative   & Positive   & Correlation \\
  Parameter     &  value    &  limits    &  limits    & coefficient~ \\ \hline
   $h_{1}^{\Zo}$  &  $-0.17$  &  $-0.72$   &   0.45     & 0.94~~~~  \\
   $h_{2}^{\Zo}$  &  $-0.09$  &  $-0.49$   &   0.36     & \\
\hline
   $h_{3}^{\Zo}$  &  $-0.01$  &  $-0.76$   &   0.67     & 0.95~~~~  \\
   $h_{4}^{\Zo}$  &  $ 0.00$  &  $-0.49$   &   0.49     & \\
\hline
   $h_{1}^\gamma$ &  $-0.07$  &  $-0.36$   &   0.24     & 0.88~~~~  \\
   $h_{2}^\gamma$ &  $-0.04$  &  $-0.25$   &   0.17     & \\
\hline
   $h_{3}^\gamma$ &  $ 0.03$  &  $-0.36$   &   0.26     & 0.96~~~~  \\
   $h_{4}^\gamma$ &  $-0.00$  &  $-0.23$   &   0.25     & \\
\hline
\end{tabular}
\icaption{Results for two-dimensional fits at $\Delta \log(Likelihood) = 3$. In each fit 
          all other six parameters are kept at zero.
\label{tab:h_corr}}
\end{table}

The main sources of systematic uncertainties are considered. The influence 
of the angular resolution for jets and photons is found to be negligible. The 
uncertainty in the trigger efficiency for the \EENNG{} channel has a negligible 
impact on the limits. The systematic error due to the limited Monte Carlo 
statistics is computed to be 0.02 for limits on individual couplings and 0.03 
for two-dimensional limits. The systematic uncertainty coming from the Monte 
Carlo modeling of detector inefficiencies and the resolution in the photon 
energy is estimated to be below 0.02. These effects 
are included in the limits presented.

If the data are interpreted in terms of new physics scales using formulae
(\ref{for:lambda13}) and (\ref{for:lambda24}), the following 
limits (95\% CL) to the new physics scales are obtained: 

\begin{center}
\begin{tabular}{lcl}
$\Lambda_{1\Zo}    > \phantom{0}867 \GeV$ & ~~ & $\Lambda_{1\gamma} > \phantom{0}947 \GeV $ \\
$\Lambda_{2\Zo}    > \phantom{0}270 \GeV$ & ~~ & $\Lambda_{2\gamma} > \phantom{0}288 \GeV $ \\
$\Lambda_{3\Zo}    > \phantom{0}652 \GeV$ & ~~ & $\Lambda_{3\gamma} >           1350 \GeV $ \\
$\Lambda_{4\Zo}    > \phantom{0}256 \GeV$ & ~~ & $\Lambda_{4\gamma} > \phantom{0}309 \GeV $. \\
\end{tabular}
\end{center}

\noindent
To obtain these results, the maximum of the likelihood is taken at the
Standard Model expectation. To determine the confidence levels the probability 
distributions are normalized over the physically allowed range of the parameters.

%
%
\section*{Acknowledgements}
\indent

We wish to express our gratitude to the CERN accelerator divisions
for the excellent performance of the LEP machine. We acknowledge
the effort of the engineers and technicians who have participated
in the construction and maintenance of the experiment.

%
%
\bibliographystyle{l3style}

%
%
\newpage
\typeout{   }     
\typeout{Using author list for paper 208 -- ? }
\typeout{$Modified: Tue May  2 13:45:26 2000 by clare $}
\typeout{!!!!  This should only be used with document option a4p!!!!}
\typeout{   }
%
%
%
%
%
%

\newcount\tutecount  \tutecount=0
\def\tutenum#1{\global\advance\tutecount by 1 \xdef#1{\the\tutecount}}
\def\tute#1{$^{#1}$}
\tutenum\aachen            
\tutenum\nikhef            
\tutenum\mich              
\tutenum\lapp              
\tutenum\basel             
\tutenum\lsu               
\tutenum\beijing           
\tutenum\berlin            
\tutenum\bologna           
\tutenum\tata              
\tutenum\ne                
\tutenum\bucharest         
\tutenum\budapest          
\tutenum\mit               
\tutenum\debrecen          
\tutenum\florence          
\tutenum\cern              
\tutenum\wl                
\tutenum\geneva            
\tutenum\hefei             
\tutenum\seft              
\tutenum\lausanne          
\tutenum\lecce             
\tutenum\lyon              
\tutenum\madrid            
\tutenum\milan             
\tutenum\moscow            
\tutenum\naples            
\tutenum\cyprus            
\tutenum\nymegen           
\tutenum\caltech           
\tutenum\perugia           
\tutenum\cmu               
\tutenum\prince            
\tutenum\rome              
\tutenum\peters            
\tutenum\potenza           
\tutenum\salerno           
\tutenum\ucsd              
\tutenum\santiago          
\tutenum\sofia             
\tutenum\korea             
\tutenum\alabama           
\tutenum\utrecht           
\tutenum\purdue            
\tutenum\psinst            
\tutenum\zeuthen           
\tutenum\eth               
\tutenum\hamburg           
\tutenum\taiwan            
\tutenum\tsinghua          

{
\parskip=0pt
\noindent
{\bf The L3 Collaboration:}
\ifx\selectfont\undefined
 \baselineskip=10.8pt
 \baselineskip\baselinestretch\baselineskip
 \normalbaselineskip\baselineskip
 \ixpt
\else
 \fontsize{9}{10.8pt}\selectfont
\fi
\medskip
\tolerance=10000
\hbadness=5000
\raggedright
\hsize=162truemm\hoffset=0mm
\def\r{\rlap,}
\noindent

M.Acciarri\r\tute\milan\
P.Achard\r\tute\geneva\ 
O.Adriani\r\tute{\florence}\ 
M.Aguilar-Benitez\r\tute\madrid\ 
J.Alcaraz\r\tute\madrid\ 
G.Alemanni\r\tute\lausanne\
J.Allaby\r\tute\cern\
A.Aloisio\r\tute\naples\ 
M.G.Alviggi\r\tute\naples\
G.Ambrosi\r\tute\geneva\
% ...
\newpage
\begin{list}{A}{\itemsep=0pt plus 0pt minus 0pt\parsep=0pt plus 0pt minus 0pt
                \topsep=0pt plus 0pt minus 0pt}
\item[\aachen]
 I. Physikalisches Institut, RWTH, D-52056 Aachen, FRG$^{\S}$\\
 III. Physikalisches Institut, RWTH, D-52056 Aachen, FRG$^{\S}$
\item[\nikhef] National Institute for High Energy Physics, NIKHEF, 
     and University of Amsterdam, NL-1009 DB Amsterdam, The Netherlands
\item[\mich] University of Michigan, Ann Arbor, MI 48109, USA
\item[\lapp] Laboratoire d'Annecy-le-Vieux de Physique des Particules, 
     LAPP,IN2P3-CNRS, BP 110, F-74941 Annecy-le-Vieux CEDEX, France
\item[\basel] Institute of Physics, University of Basel, CH-4056 Basel,
     Switzerland
\item[\lsu] Louisiana State University, Baton Rouge, LA 70803, USA
\item[\beijing] Institute of High Energy Physics, IHEP, 
  100039 Beijing, China$^{\triangle}$ 
\item[\berlin] Humboldt University, D-10099 Berlin, FRG$^{\S}$
\item[\bologna] University of Bologna and INFN-Sezione di Bologna, 
     I-40126 Bologna, Italy
\item[\tata] Tata Institute of Fundamental Research, Bombay 400 005, India
\item[\ne] Northeastern University, Boston, MA 02115, USA
\item[\bucharest] Institute of Atomic Physics and University of Bucharest,
     R-76900 Bucharest, Romania
\item[\budapest] Central Research Institute for Physics of the 
     Hungarian Academy of Sciences, H-1525 Budapest 114, Hungary$^{\ddag}$
\item[\mit] Massachusetts Institute of Technology, Cambridge, MA 02139, USA
\item[\debrecen] KLTE-ATOMKI, H-4010 Debrecen, Hungary$^\P$
\item[\florence] INFN Sezione di Firenze and University of Florence, 
     I-50125 Florence, Italy
\item[\cern] European Laboratory for Particle Physics, CERN, 
     CH-1211 Geneva 23, Switzerland
\item[\wl] World Laboratory, FBLJA  Project, CH-1211 Geneva 23, Switzerland
\item[\geneva] University of Geneva, CH-1211 Geneva 4, Switzerland
\item[\hefei] Chinese University of Science and Technology, USTC,
      Hefei, Anhui 230 029, China$^{\triangle}$
\item[\seft] SEFT, Research Institute for High Energy Physics, P.O. Box 9,
      SF-00014 Helsinki, Finland
\item[\lausanne] University of Lausanne, CH-1015 Lausanne, Switzerland
\item[\lecce] INFN-Sezione di Lecce and Universit\'a Degli Studi di Lecce,
     I-73100 Lecce, Italy
\item[\lyon] Institut de Physique Nucl\'eaire de Lyon, 
     IN2P3-CNRS,Universit\'e Claude Bernard, 
     F-69622 Villeurbanne, France
\item[\madrid] Centro de Investigaciones Energ{\'e}ticas, 
     Medioambientales y Tecnolog{\'\i}cas, CIEMAT, E-28040 Madrid,
     Spain${\flat}$ 
\item[\milan] INFN-Sezione di Milano, I-20133 Milan, Italy
\item[\moscow] Institute of Theoretical and Experimental Physics, ITEP, 
     Moscow, Russia
\item[\naples] INFN-Sezione di Napoli and University of Naples, 
     I-80125 Naples, Italy
\item[\cyprus] Department of Natural Sciences, University of Cyprus,
     Nicosia, Cyprus
\item[\nymegen] University of Nijmegen and NIKHEF, 
     NL-6525 ED Nijmegen, The Netherlands
\item[\caltech] California Institute of Technology, Pasadena, CA 91125, USA
\item[\perugia] INFN-Sezione di Perugia and Universit\'a Degli 
     Studi di Perugia, I-06100 Perugia, Italy   
\item[\cmu] Carnegie Mellon University, Pittsburgh, PA 15213, USA
\item[\prince] Princeton University, Princeton, NJ 08544, USA
\item[\rome] INFN-Sezione di Roma and University of Rome, ``La Sapienza",
     I-00185 Rome, Italy
\item[\peters] Nuclear Physics Institute, St. Petersburg, Russia
\item[\potenza] INFN-Sezione di Napoli and University of Potenza, 
     I-85100 Potenza, Italy
\item[\salerno] University and INFN, Salerno, I-84100 Salerno, Italy
\item[\ucsd] University of California, San Diego, CA 92093, USA
\item[\santiago] Dept. de Fisica de Particulas Elementales, Univ. de Santiago,
     E-15706 Santiago de Compostela, Spain
\item[\sofia] Bulgarian Academy of Sciences, Central Lab.~of 
     Mechatronics and Instrumentation, BU-1113 Sofia, Bulgaria
\item[\korea]  Laboratory of High Energy Physics, 
     Kyungpook National University, 702-701 Taegu, Republic of Korea
\item[\alabama] University of Alabama, Tuscaloosa, AL 35486, USA
\item[\utrecht] Utrecht University and NIKHEF, NL-3584 CB Utrecht, 
     The Netherlands
\item[\purdue] Purdue University, West Lafayette, IN 47907, USA
\item[\psinst] Paul Scherrer Institut, PSI, CH-5232 Villigen, Switzerland
\item[\zeuthen] DESY, D-15738 Zeuthen, 
     FRG
\item[\eth] Eidgen\"ossische Technische Hochschule, ETH Z\"urich,
     CH-8093 Z\"urich, Switzerland
\item[\hamburg] University of Hamburg, D-22761 Hamburg, FRG
\item[\taiwan] National Central University, Chung-Li, Taiwan, China
\item[\tsinghua] Department of Physics, National Tsing Hua University,
      Taiwan, China
\item[\S]  Supported by the German Bundesministerium 
        f\"ur Bildung, Wissenschaft, Forschung und Technologie
\item[\ddag] Supported by the Hungarian OTKA fund under contract
numbers T019181, F023259 and T024011.
\item[\P] Also supported by the Hungarian OTKA fund under contract
  numbers T22238 and T026178.
\item[$\flat$] Supported also by the Comisi\'on Interministerial de Ciencia y 
        Tecnolog{\'\i}a.
\item[$\sharp$] Also supported by CONICET and Universidad Nacional de La Plata,
        CC 67, 1900 La Plata, Argentina.
\item[$\diamondsuit$] Also supported by Panjab University, Chandigarh-160014, 
        India.
\item[$\triangle$] Supported by the National Natural Science
  Foundation of China.
\end{list}
}
\vfill


%
%
%
\newpage
\begin{figure}
\begin{center}
\includegraphics[width=15.0truecm]{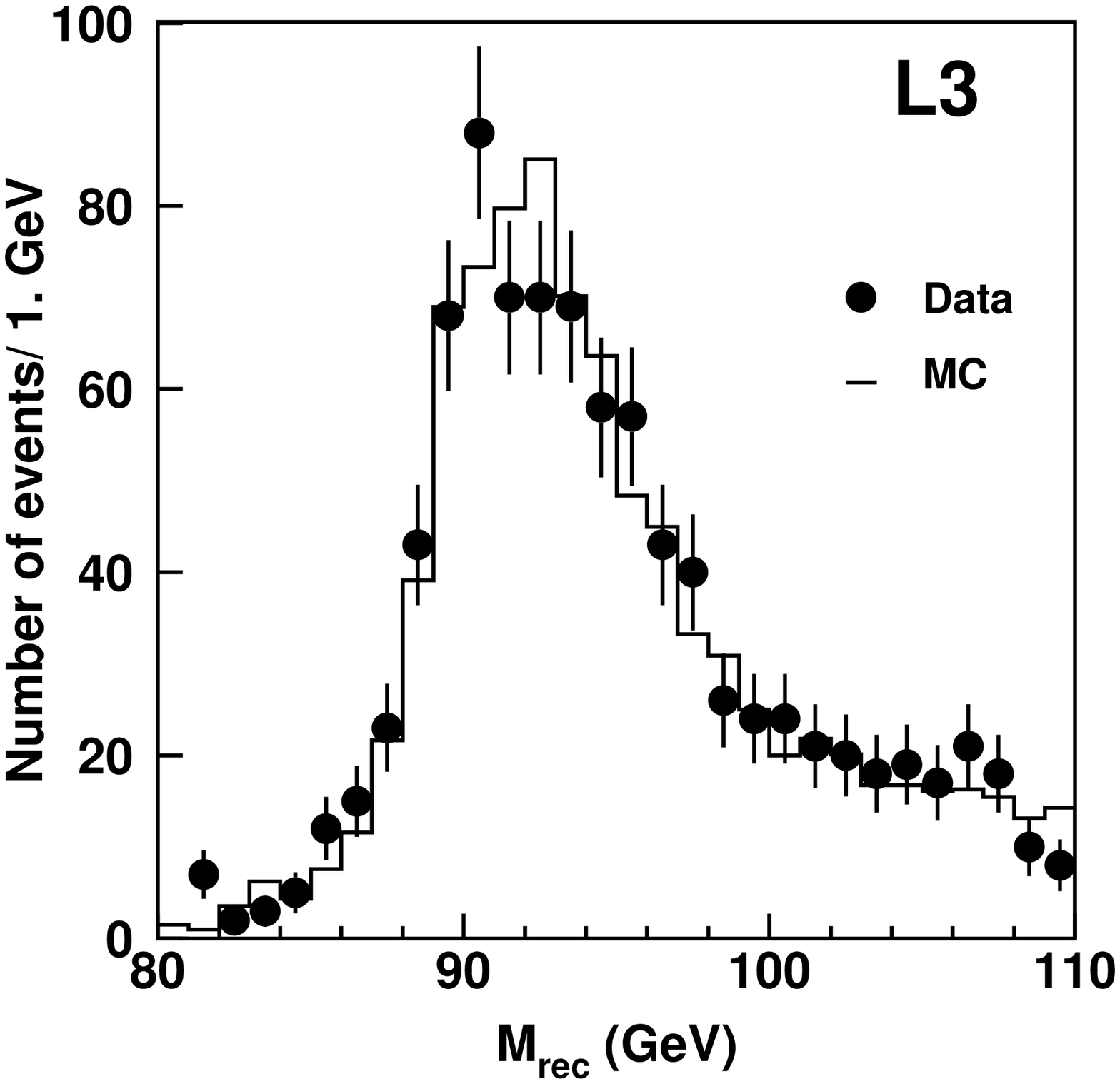}
\icaption{\label{fig:recomass_qq}
          Distribution of the  mass recoiling from the photon candidate
          in $\EEQQG$ events. The points are data and the histogram is the 
          Standard Model Monte Carlo prediction.}
\end{center}
\end{figure}
%
%
\newpage
\begin{figure}
\begin{center}
\includegraphics[width=15.0truecm]{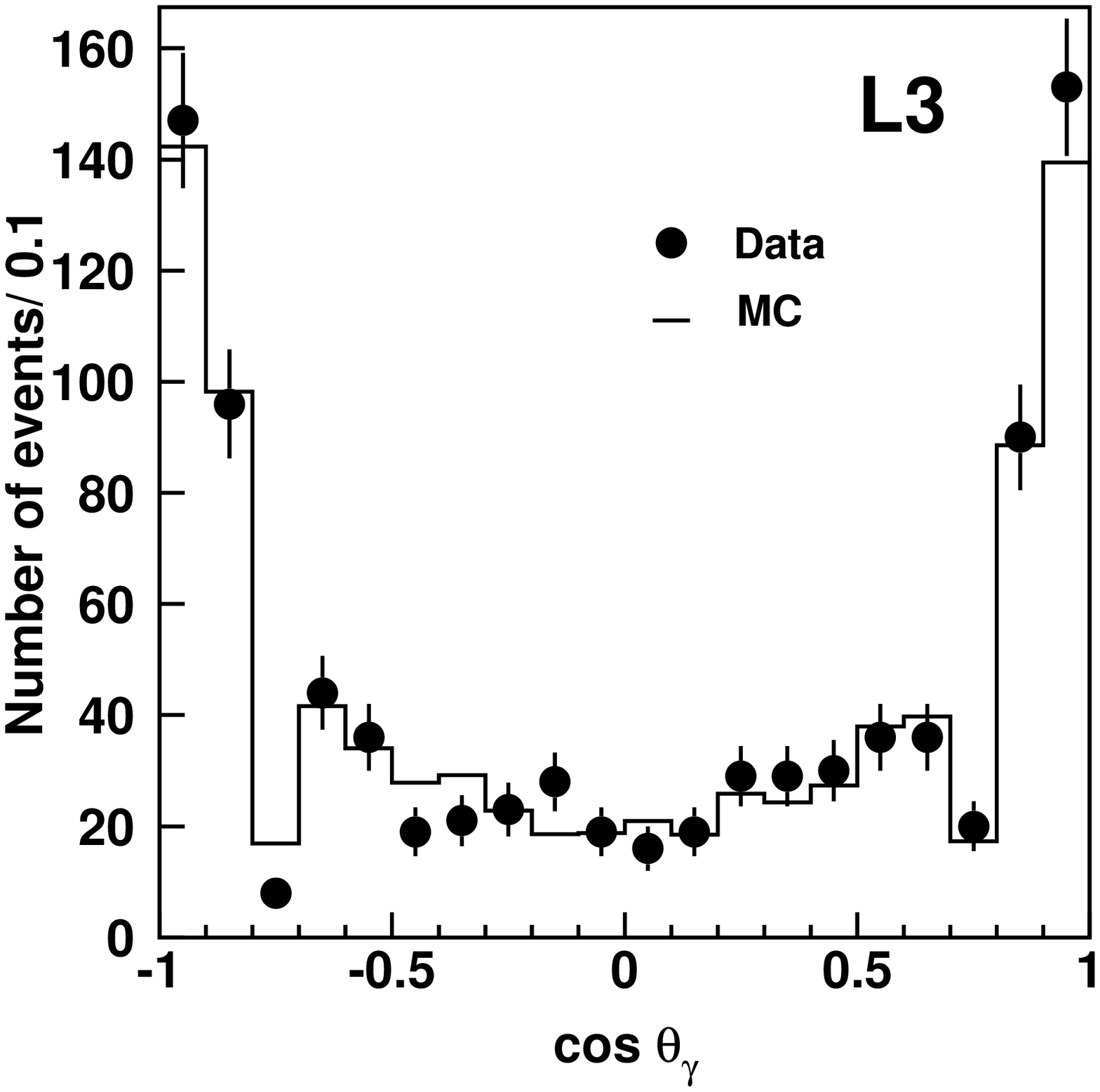}
\icaption{\label{fig:thetafo_qq}
          Polar angle distribution from the photon candidates in 
          $\EEQQG$ events. The points are data and the histogram is the
          Standard Model Monte Carlo prediction.}
\end{center}
\end{figure}
%
%
\newpage
\begin{figure}
\begin{center}
\includegraphics[width=15.0truecm]{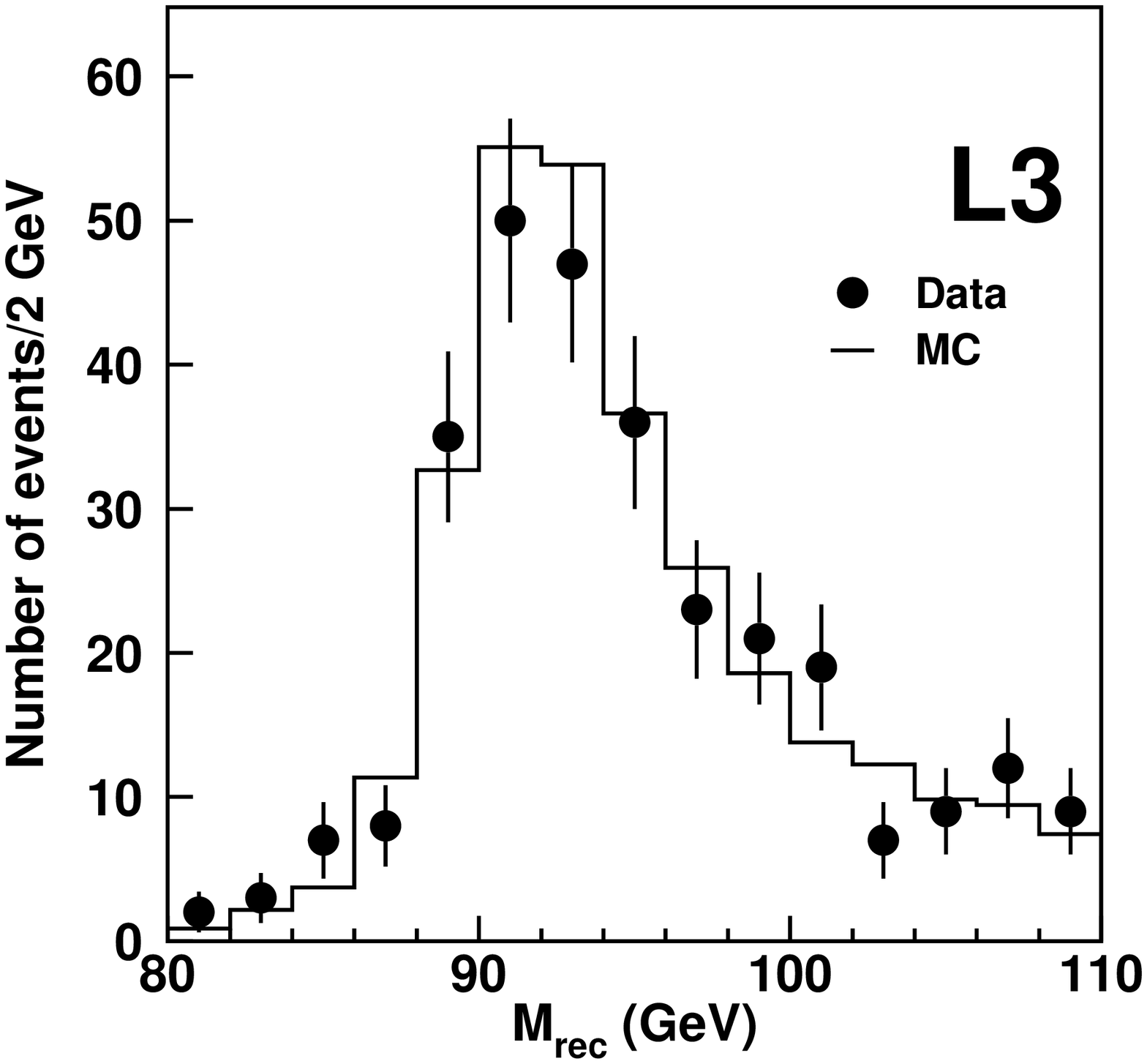}
\icaption{\label{fig:recoilmass_nunug}
          Distribution of the mass recoiling from the photon candidate
          in $\EENNG$ events. The points are data and the histogram is the 
          Standard Model Monte Carlo prediction.}
\end{center}
\end{figure}
%
%
\newpage
\begin{figure}
\begin{center}
\includegraphics[width=15.0truecm]{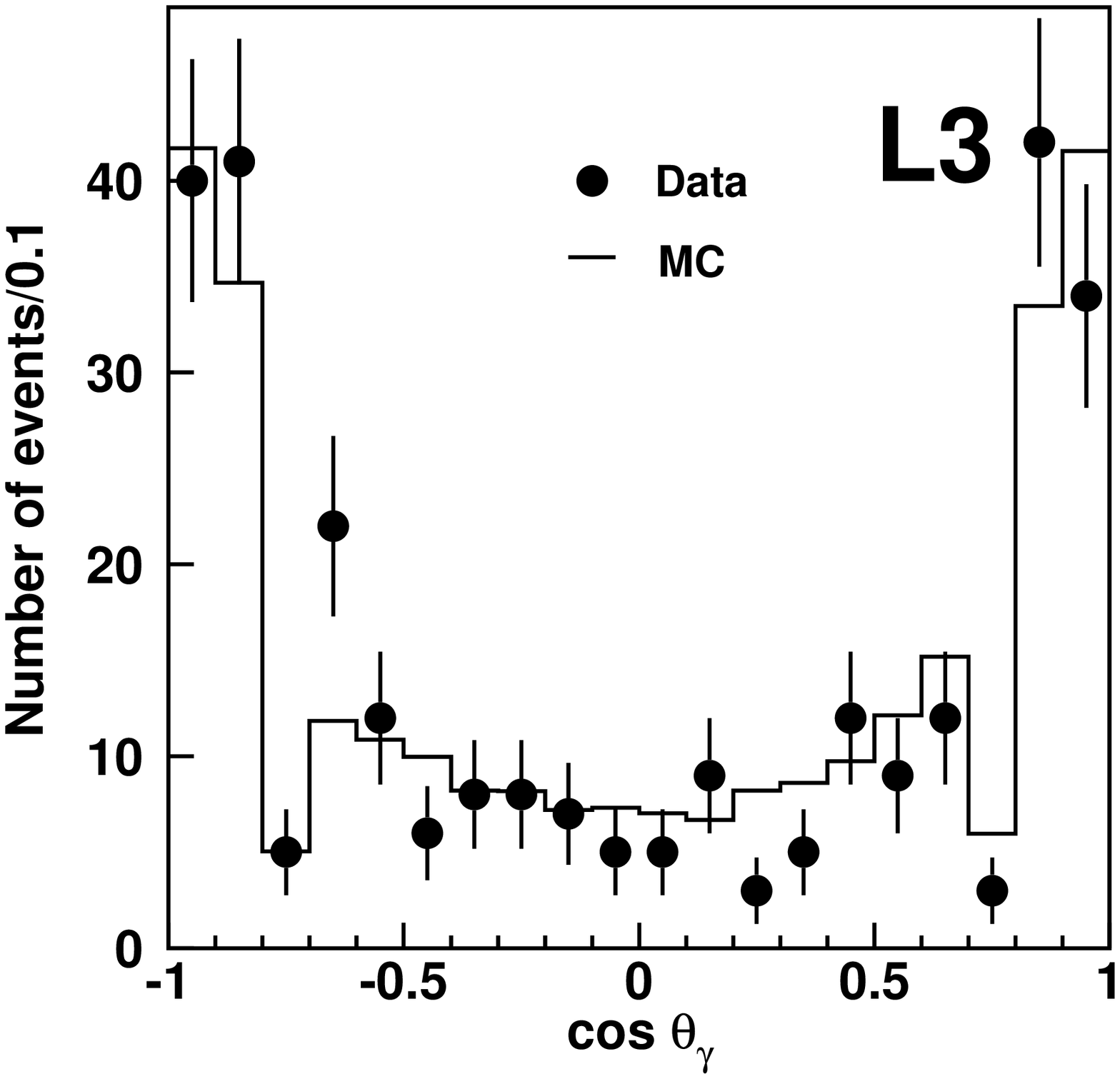}
\icaption{\label{fig:costheta_nunug}
          Polar angle distribution from the photon candidates in 
          $\EENNG$ events. The points are data and the histogram is the 
          Standard Model Monte Carlo prediction.}
\end{center}
\end{figure}
%
%
\newpage
\begin{figure}
\begin{center}
\includegraphics[width=15.0truecm]{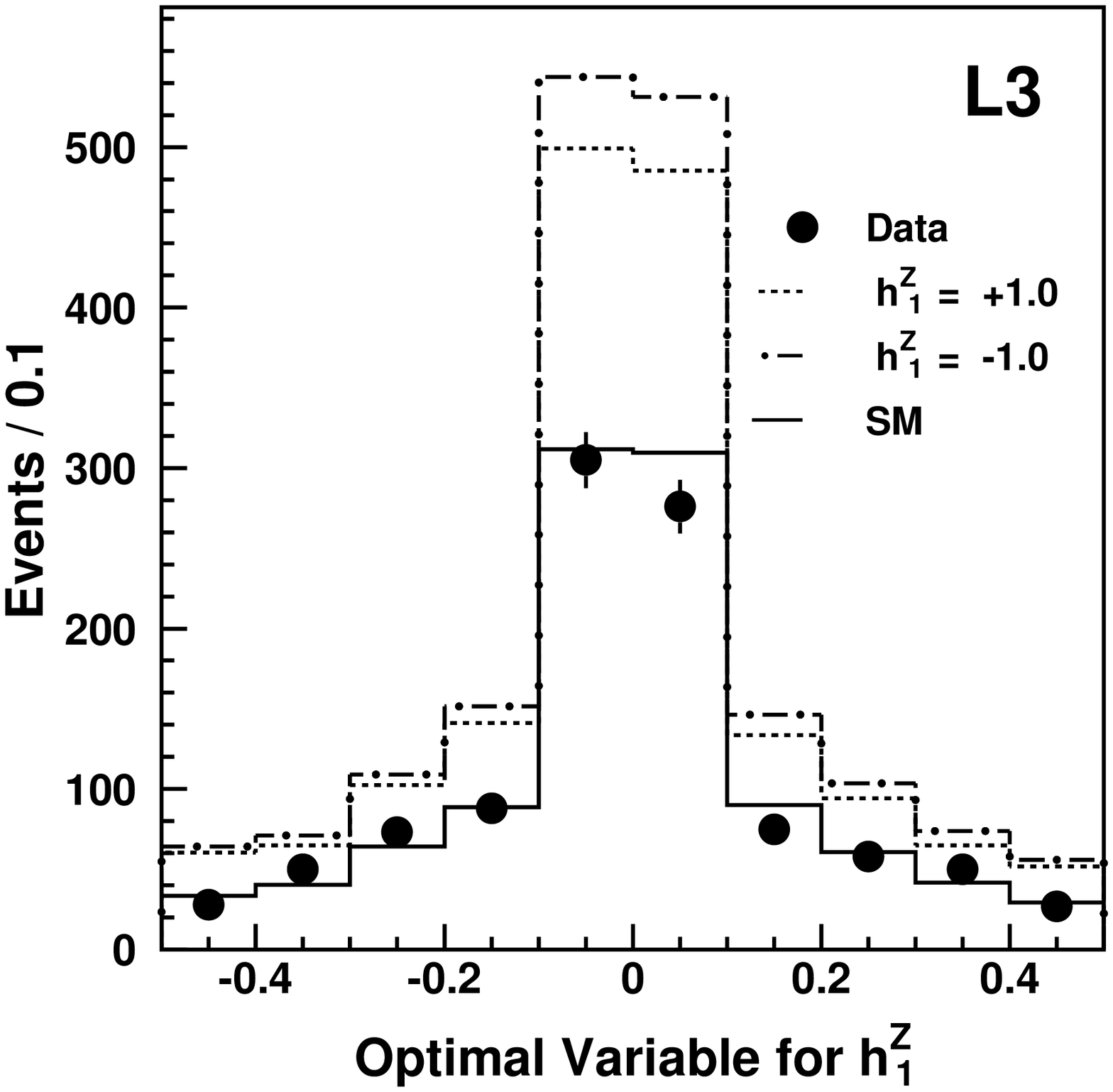}
\icaption{\label{fig:h1z_oo}
          Distribution of the optimal variable for the CP violating 
          coupling $h_{1}^{\Zo}$. Data are shown, together with the 
          expectations for the Standard Model (SM) and
          for anomalous couplings, $h_{1}^{\Zo}=\pm 1$.} 
\end{center}
\end{figure}
%
%
\newpage
\begin{figure}
\begin{center}
\includegraphics[width=15.0truecm]{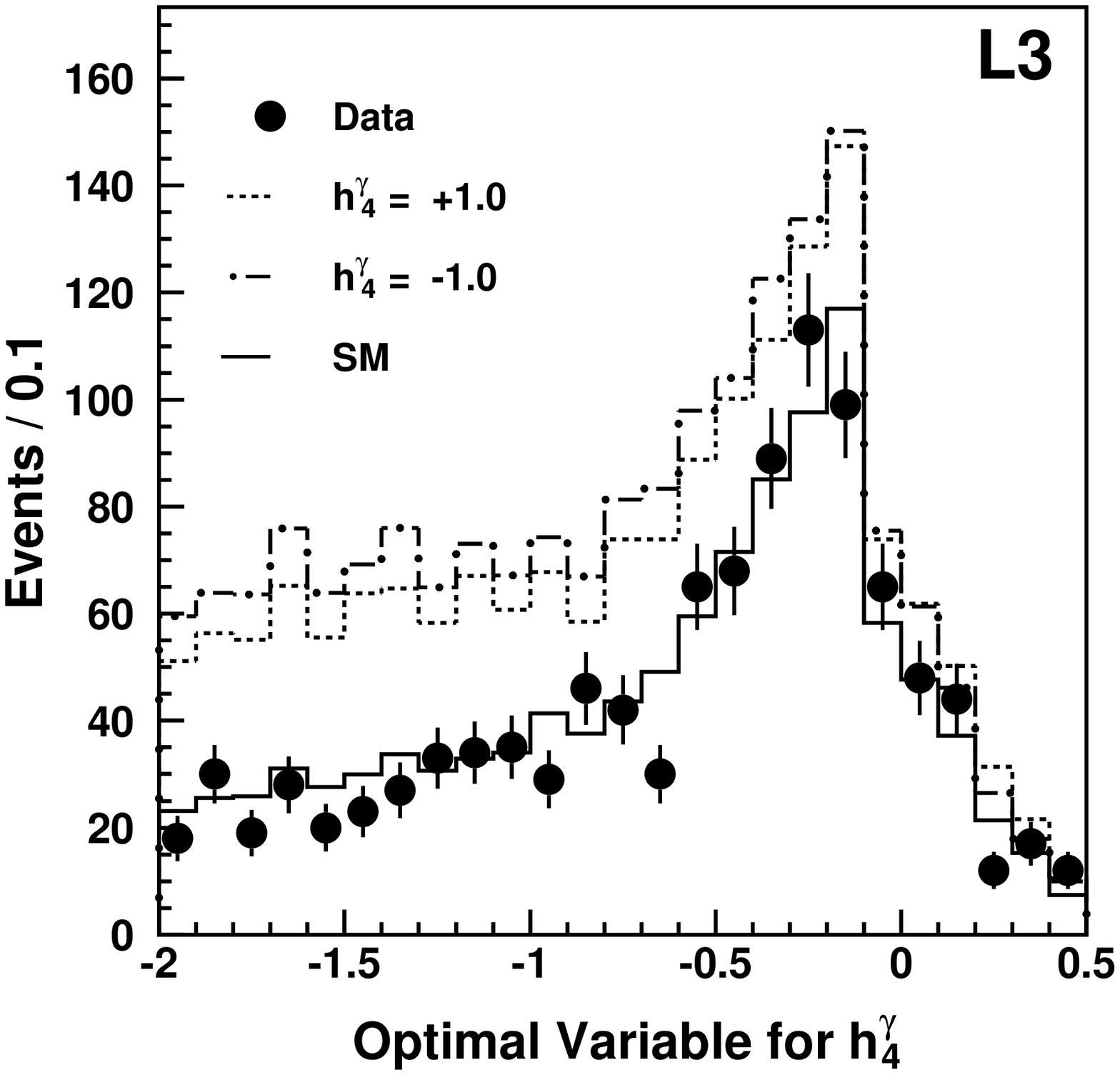}
\icaption{\label{fig:h4g_oo}
          Distribution of the optimal variable for the CP conserving 
          coupling $h_{4}^{\gamma}$. Data are shown, together with the 
          expectations for the Standard Model (SM) and
          for anomalous couplings, $h_{4}^{\gamma}=\pm 1$.} 
\end{center}
\end{figure}
%
\newpage
\begin{figure}
\begin{center}
\vskip-2.5truecm
\includegraphics[width=13.0truecm]{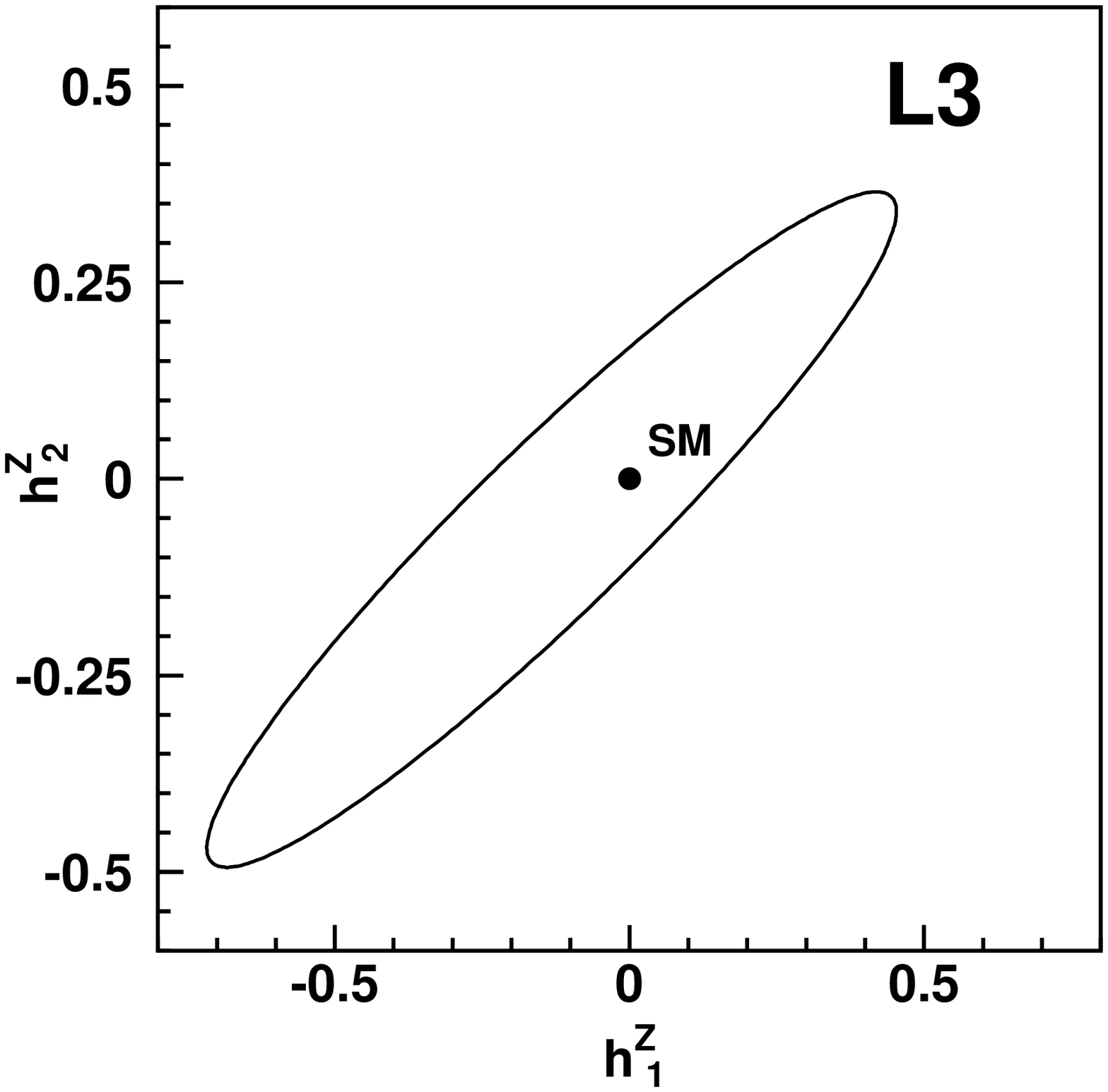}
\vskip-1.5truecm
\includegraphics[width=13.0truecm]{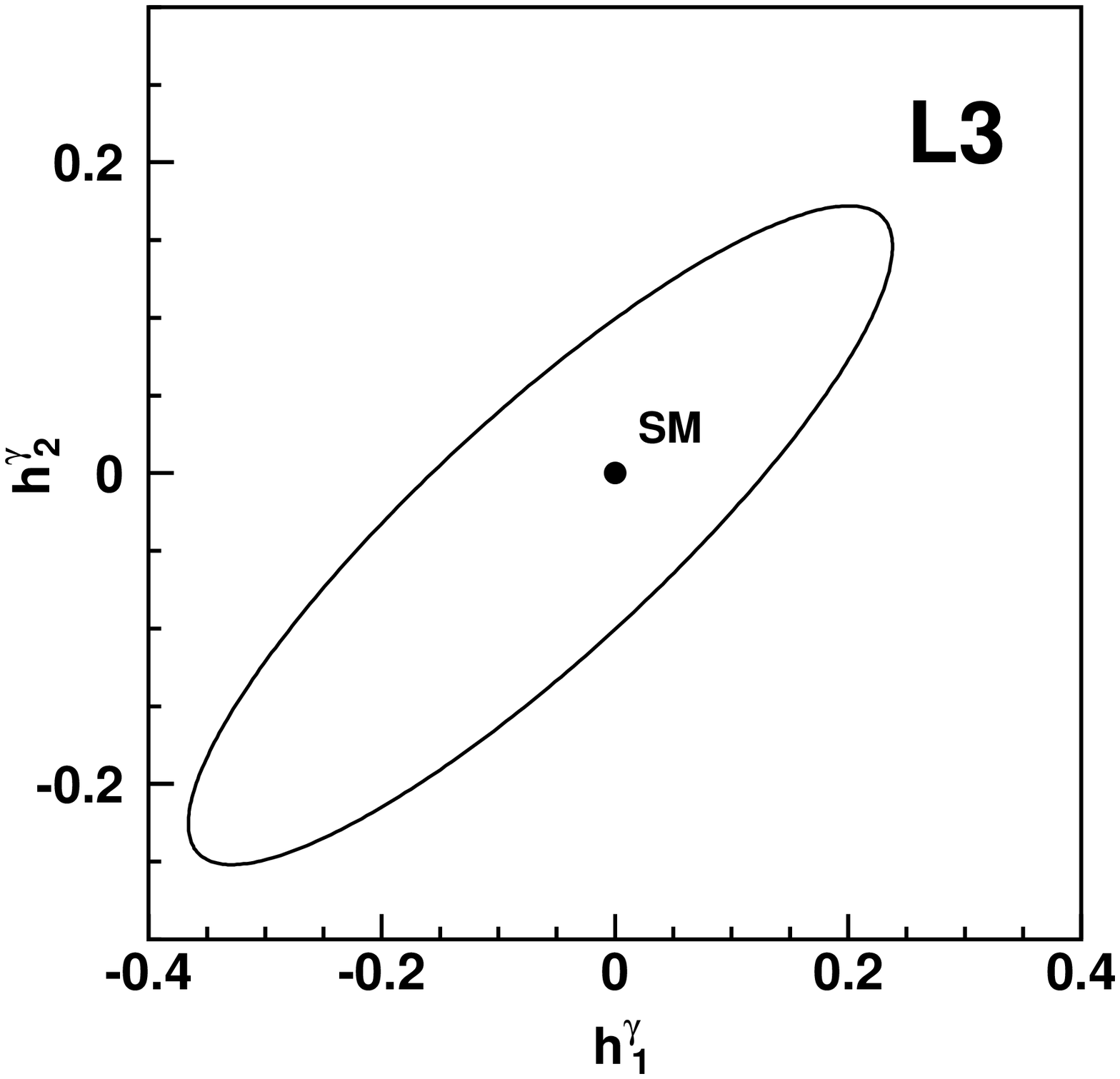}
\vspace*{0.5truecm}
\icaption{\label{fig:elipse2}
          Limits at $95\%$ CL on the CP-violating coupling parameters,
          $h_{2}^{\Zo}$ versus $h_{1}^{\Zo}$ and $h_{2}^{\gamma}$ versus
          $h_{1}^{\gamma}$. The Standard Model predictions are indicated 
          by the points. The regions outside the contours are excluded.}
\end{center}
\end{figure}
%
%
\newpage
\begin{figure}
\begin{center}
\vskip-2.5truecm
\includegraphics[width=13.0truecm]{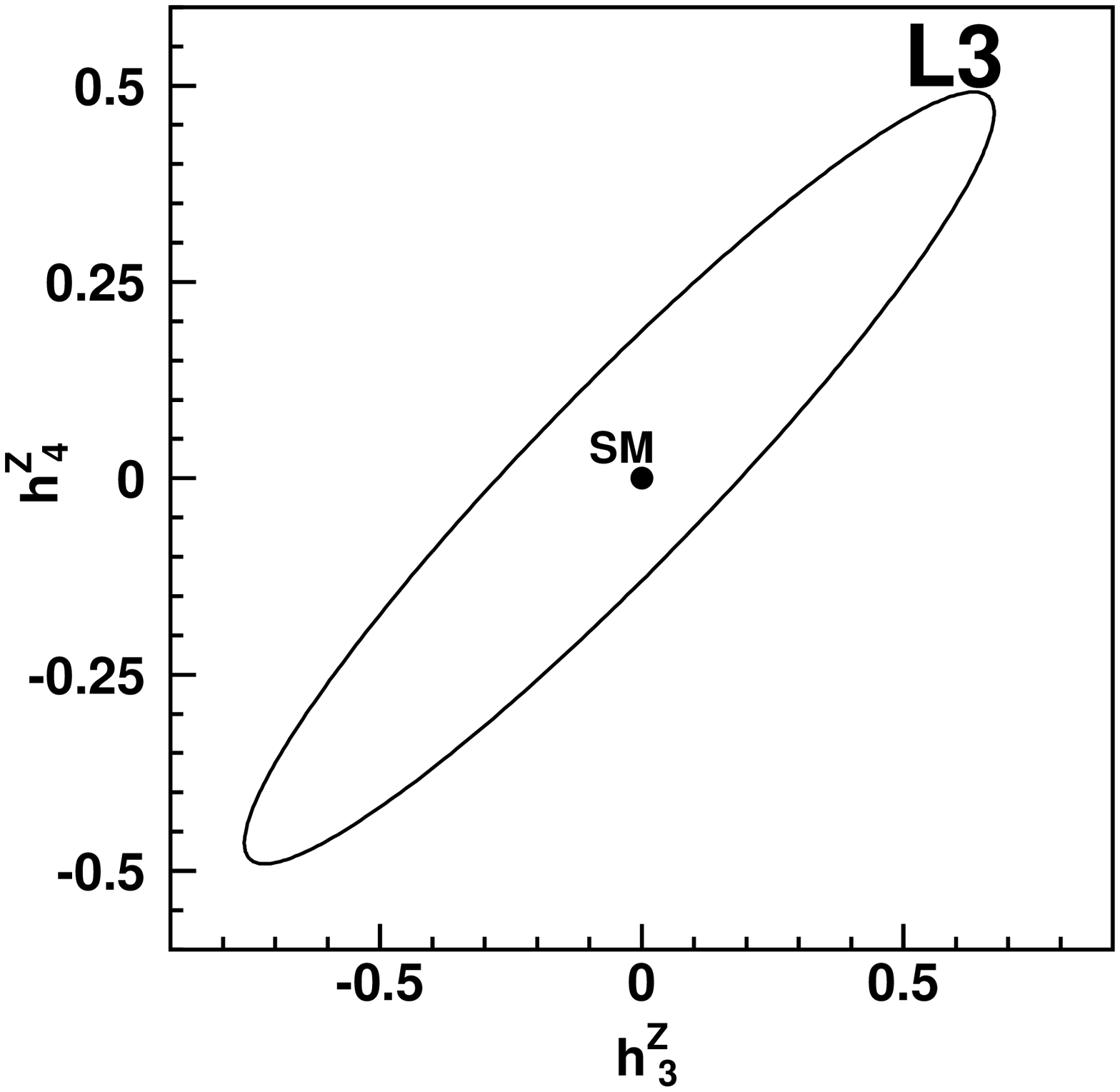}
\vskip-1.5truecm
\includegraphics[width=13.0truecm]{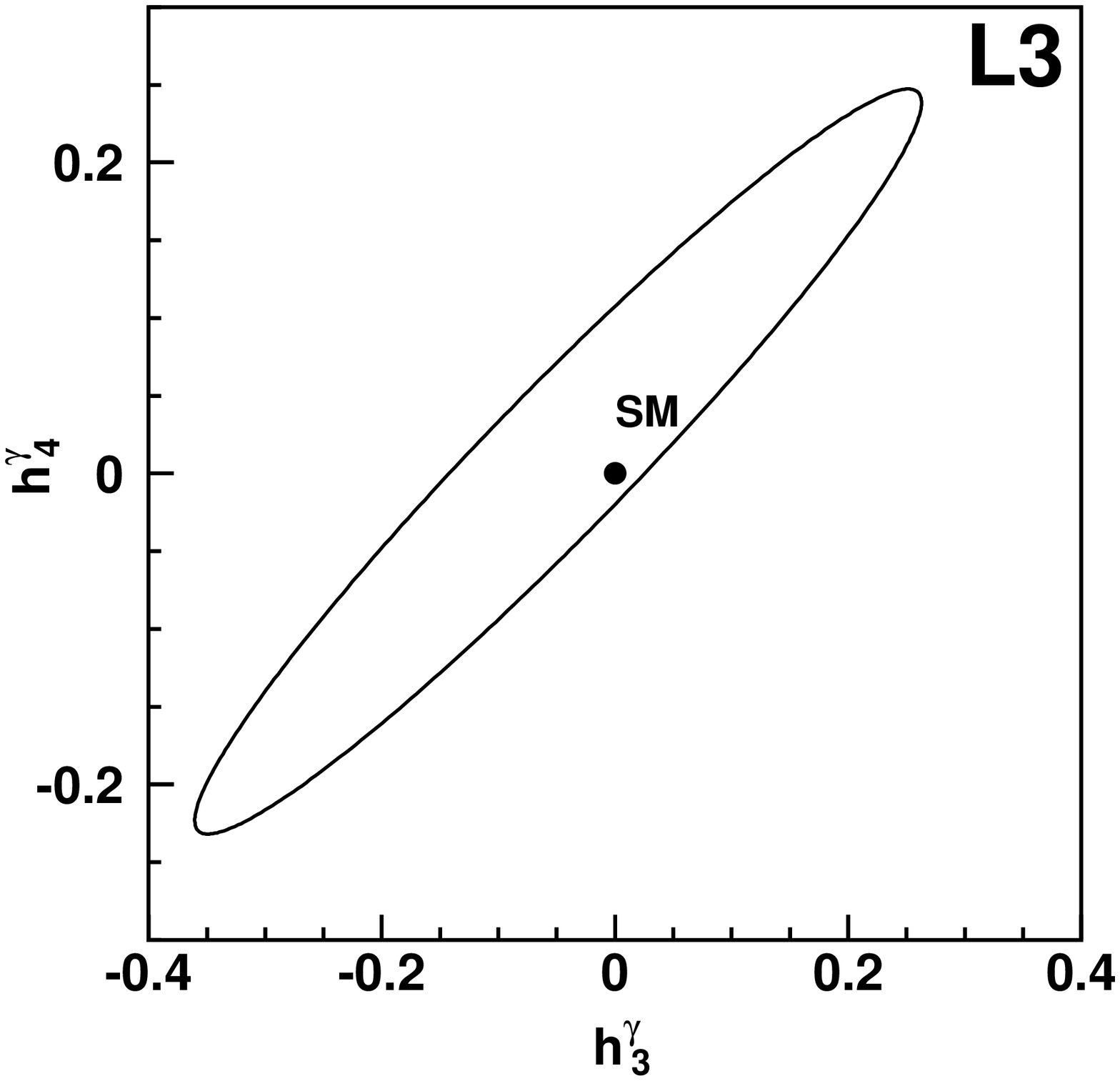}
\vspace*{0.5truecm}
\icaption{\label{fig:elipse1}
          Limits at $95\%$ CL on the CP-conserving coupling parameters,
          $h_{4}^{\Zo}$ versus $h_{3}^{\Zo}$ and $h_{4}^{\gamma}$ versus
          $h_{3}^{\gamma}$. The Standard Model predictions are indicated 
          by the points. The regions outside the contours are excluded.}
\end{center}
\end{figure}
%
%
%
\end{document}